\documentclass[a4paper]{jpconf_Ref-11}
\usepackage{citesort}
\usepackage{iopams}
\usepackage{textcomp}  
\usepackage{graphicx}

\begin{document}

\title{Ge clusters and wetting layers forming from granular films  on the Si(001) surface}

\author{M~S Storozhevykh,
L~V Arapkina
and V~A Yuryev
}

\address{
A.\,M.\,Prokhorov General Physics Institute of the Russian Academy of Sciences,\\ 38 Vavilov Street, Moscow, 119991, Russia
}
\ead{storozhevykh@kapella.gpi.ru}

\begin{abstract}
The report studies transformation of a Ge granular film deposited at room temperature on the Si(001) surface to the Ge/Si(001) heterostructure as a result of rapid heating and annealing at 600\,{\textcelsius}.
As a result of the short-term annealing at 600\,{\textcelsius} in conditions of a closed system, the Ge granular film transforms to the usual wetting layer and Ge clusters with multimodal size distribution and the Ge oval drops having the highest number density. After the long-term thermal treatment of the Ge film at the same temperature, Ge drops disappear; the large clusters increase their sizes at the expense of the smaller ones. The total density of Ge clusters on the surface drastically decreases. The wetting layer mixed $c(4\times 2)+p(2\times 2)$ reconstruction transforms to the single $c(4\times 2)$ one which is likely thermodynamically favoured. Pyramids or domes are not observed on the surface after any annealing.
 \end{abstract}

\section{\label{sec:intro}Introduction}

This paper presents an experimental study of transformation of a disordered Ge film on Si(001) to the Ge/Si(001) heterostructure. We deposited Ge at the room temperature on the Si(001) layer  grown by means of  the molecular-beam epitaxy (MBE)  on the Si substrate and explored crystallization of the obtained Ge granular film as a result of rapid heating and isothermal treatments at $600$\,\textcelsius.
This experiment has given the following results: 
First of all, we have demonstrated that the Ge/Si(001) heterosructure formed  as a result of a short-term annealing under the conditions of a closed system consists of the usual patched wetting layer and large clusters of Ge with multimodal size distribution rather than pyramids or domes which appear when a film is deposited  in a flux of Ge atoms arriving on its surface \cite{Stability-Anealing,Domes_first,Pyramid_to_dome,Ross,Ge_Shapes}.
Ge oval drops with the lateral dimensions of about a hundred nanometers have the highest number density among the detected clusters.
Then, we detected a mixture of $c(4\times 2)$ and  $p(2\times 2)$ reconstructions on the surface of the formed wetting layer whereas the simultaneous presence of both these structures in comparable proportions on wetting layer patches is a distinctive feature of the low-temperature mode of the wetting layer growth (at $T_{\rm gr}<600$\,\textcelsius) in the MBE process \cite{initial_phase,CMOS-compatible-EMRS}.
And finally,  we have shown that the Ge drops disappear from the surface as a result of long-term isothermal annealing of the original Ge film under the conditions of an isolated system;  the sizes of the large clusters increase at the expense of the smaller ones and the total density of Ge clusters on the surface decreases by several orders of magnitude.
The wetting layer retains all main features of the one forming as a result of the short-term annealing except for the reconstruction of its patches. The latter changes to pure $c(4\times 2)$ one which is a characteristic feature of the wetting layers grown by MBE at high temperature (at $T_{\rm gr}>600$\,\textcelsius) \cite{Nucleation_high-temperatures}.



\section{\label{sec:exp}Details of experiments}

\subsection{\label{sec:samples}Samples}

Ge films ($h_{\rm Ge}= 7$\,{\AA}) were deposited from molecular beams at the room temperature (${d}h_{\rm Ge}/{d}t \approx 0.15$\,\AA/s) on Si buffers (1000\,{\AA} thick) grown in the same cycles on commercial (100)-oriented wafers of Si ($\rho$ = 12\,\textohm cm) at the temperature of 650\,\textcelsius. 
Details of the pre-growth treatments of Si wafers, which included wet chemical etching and oxide removal by short high-temperature annealing ($T_{\rm an}\sim$~900\,\textcelsius), can be found in our previous articles \cite{our_Si(001)_en,stm-rheed-EMRS,phase_transition,hydrogenation_YUR}. 
 After Ge deposition, the samples were heated to 600{\,\textcelsius} at the maximum rate achievable for the used infrared heaters (0.24{\,\textcelsius}/s), 
annealed at this temperature for 5 or 125 minutes and cooled to the room temperature at the quenching mode  (0.4\,\textcelsius/s) \cite{stm-rheed-EMRS,phase_transition}.  

\subsection{\label{sec:equipment}Techniques and equipment}

The experiments were carried out using an ultrahigh-vacuum (UHV) MBE chamber (Riber EVA~32) connected with a UHV scanning tunnelling microscope (STM) chamber (GPI~300) \cite{CMOS-compatible-EMRS,VCIAN2011,STM_GPI-Proc}. 
The rates of Ge and Si deposition and the coverages of Ge and Si ($h_{\rm Ge}$, $h_{\rm Si}$) were measured by a graduated in advance film thickness monitors (Inficon Leybold-Heraeus XTC 751-001-G1) with quartz sensors installed in the MBE chamber. 
During annealing, samples were heated from the rear side by radiators of tantalum. 
The temperature was monitored with a tungsten-rhenium thermocouple  mounted in the vacuum near the rear side of the samples and {\it in situ} graduated against  a specialized pyrometer (IMPAC IS 12-Si).
The atmosphere composition in the MBE chamber was monitored using a mass-spectrometer  residual gas analyzer (SRS RGA-200) before and during the processes. 
STM images were obtained at room temperature; they were processed using the WSxM software \cite{WSxM}.
Microphotographs of the sample surfaces were obtained with the Axiotech~100 microscope (Carl Zeiss).
Additional details concerning the used equipment can be found, e.\,g., in Refs.\,\cite{classification,CMOS-compatible-EMRS,VCIAN2011}.

\begin{figure}[t]
\includegraphics[scale=1.2]{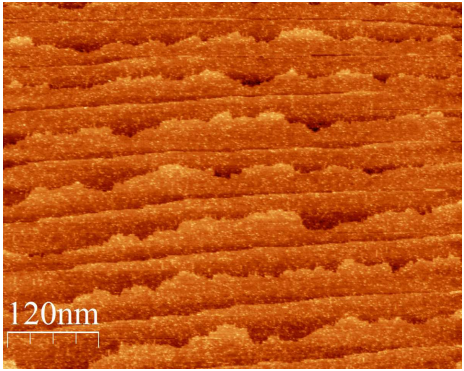}\hspace{1.5pc}%
\begin{minipage}[b]{22pc}
\caption{\label{fig:Si-STM}
STM image of the Si buffer surface   before Ge deposition ($h_{\rm Si}$ = 1000\,{\AA}, $T_{\rm gr}$ = 650\,{\textcelsius}).
 }
\end{minipage}
\end{figure}  

\section{\label{sec:results}Results and discussion}

\subsection{\label{sec:Si}Si buffer layer}

An STM image of the Si(001) surface of the buffer layer before depositing a  Ge film is demonstrated in figure \ref{fig:Si-STM}.
Alternate wide terraces ending by ${\rm S_a}$ and ${\rm S_b}$ steps \cite{Chadi-old,Misbah} with smooth and jagged edges are observed on the surface as well as numerous Si adatom clusters that is characteristic to the Si(001) surfaces cooled at the quenching mode \cite{stm-rheed-EMRS,phase_transition,SPIE_Si-Si(001)_defects}.

\begin{figure}[t]
\centering
\includegraphics[scale=.7]{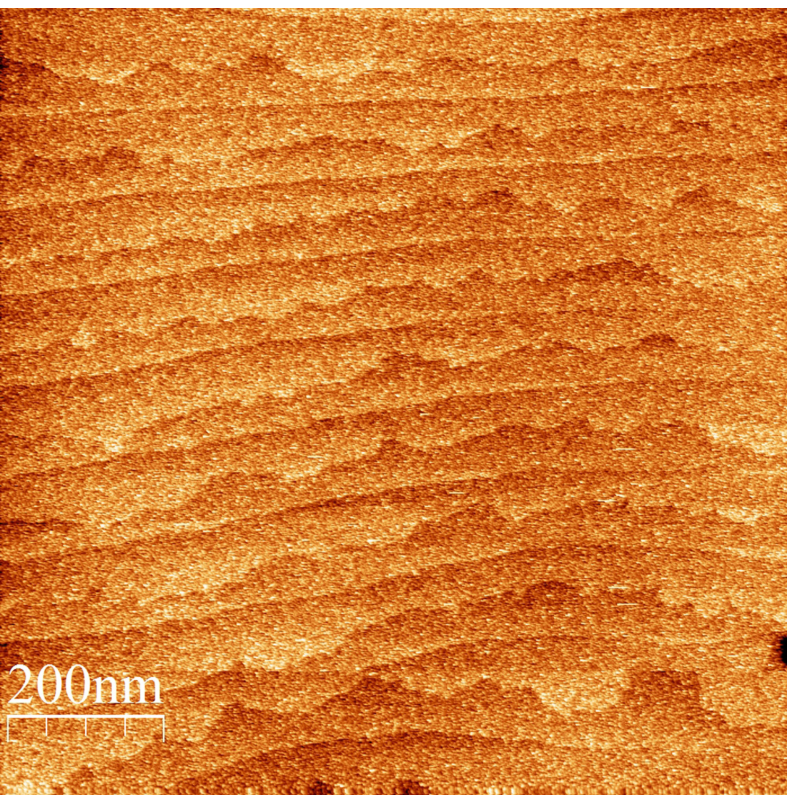}(a)
\includegraphics[scale=.7]{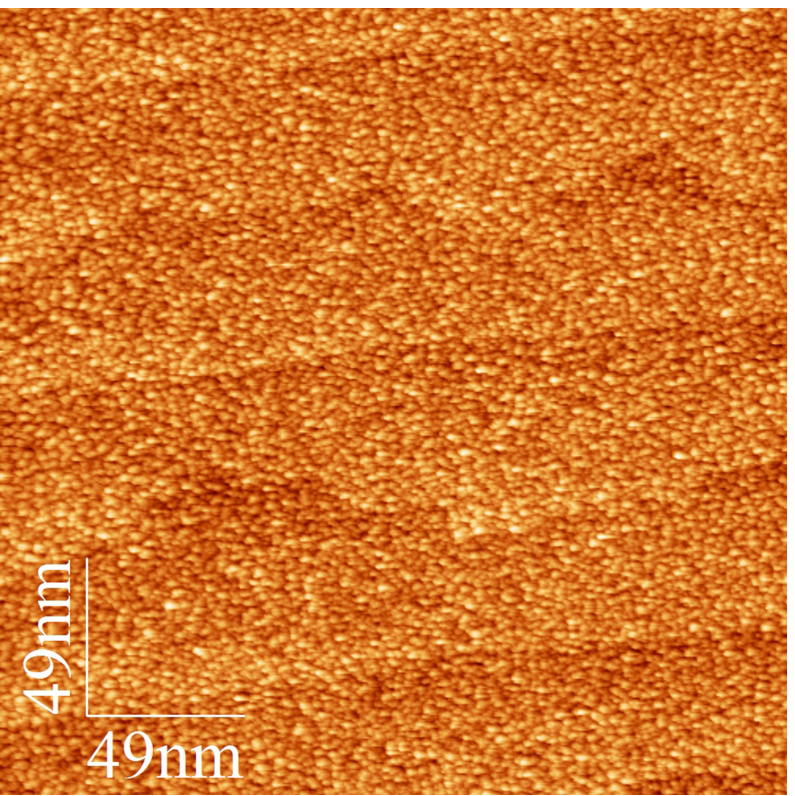}(b)\\
\includegraphics[scale=.5]{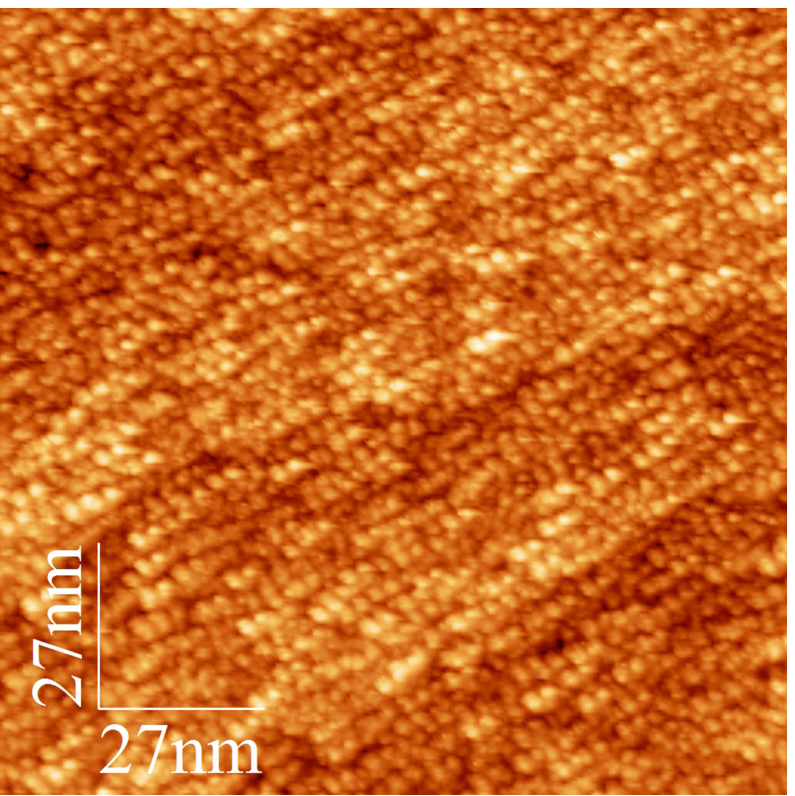}(c)
\includegraphics[scale=.5]{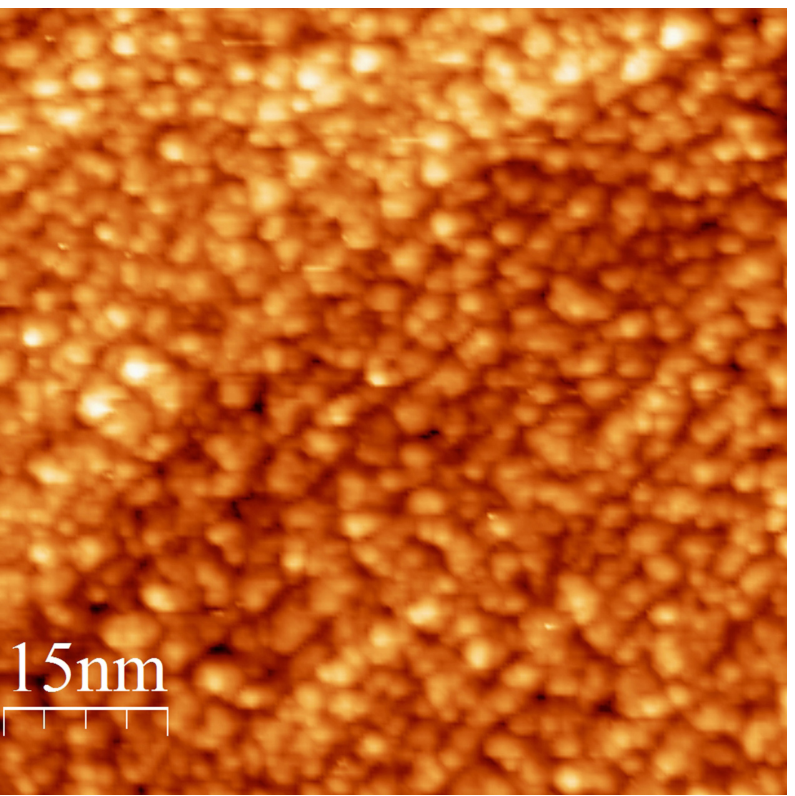}(d)
\includegraphics[scale=.5]{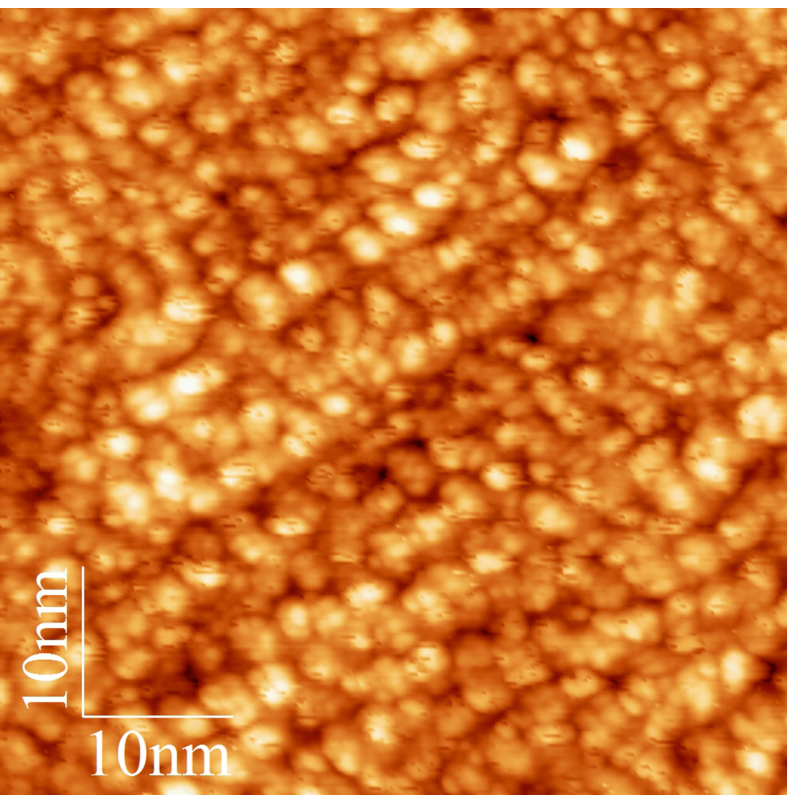}(e)
\caption{\label{fig:poly-Ge}
STM images of the sample surface after Ge deposition ($h_{\rm Ge}$ = 7\,{\AA}) at room temperature on Si(001): (a,\,b) a general view and Si terraces under the Ge film, (c,\,d,\,e) grains of the Ge film.
 }
\end{figure}

\begin{figure}[ht]
\begin{minipage}[b]{25pc}\centering
\includegraphics[scale=.55]{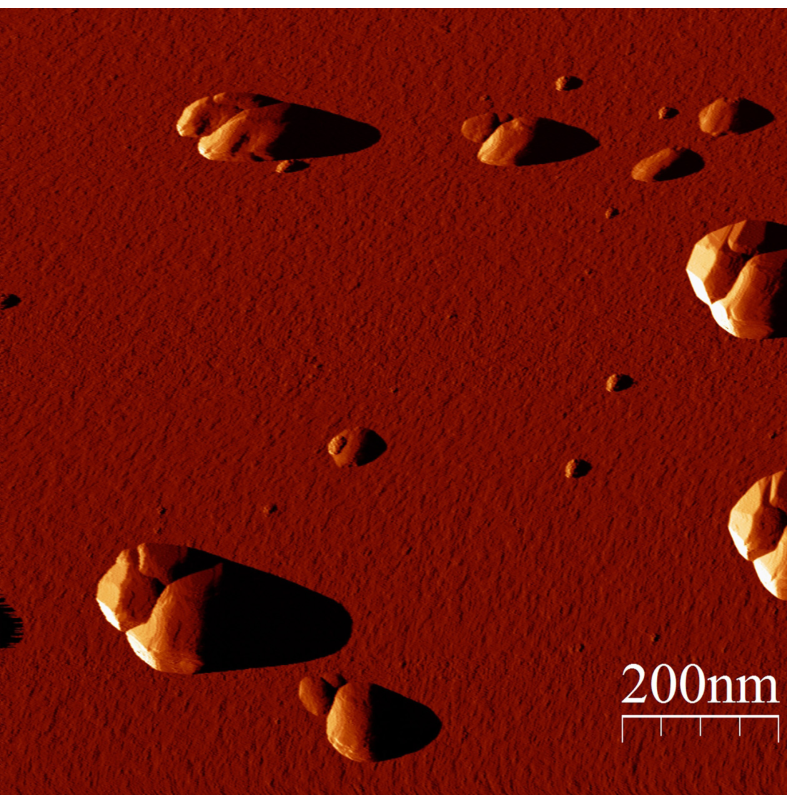}(a)
\includegraphics[scale=.55]{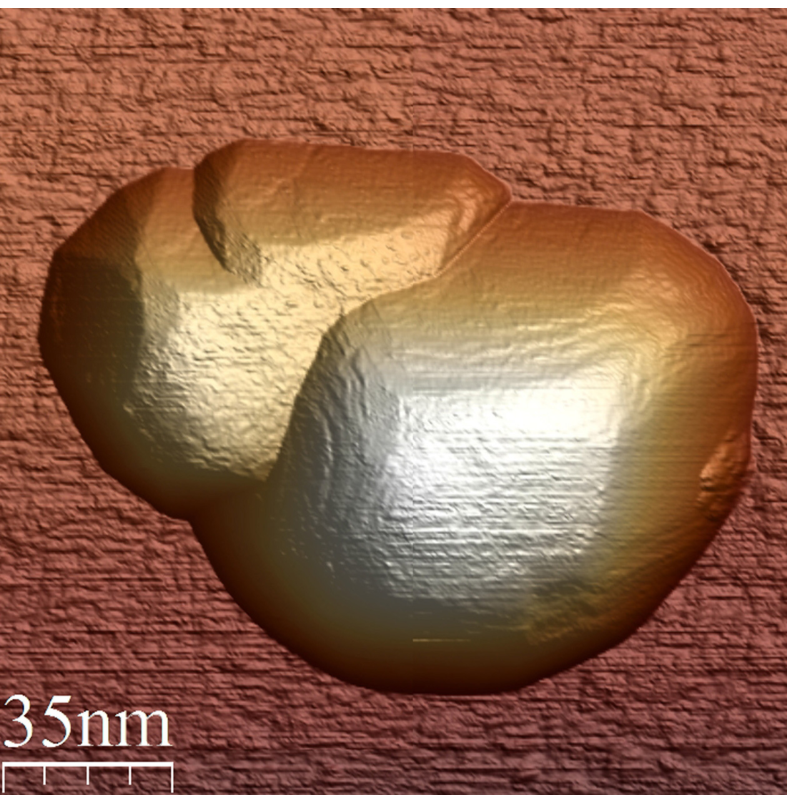}(b)\\
~\\
\includegraphics[scale=.47]{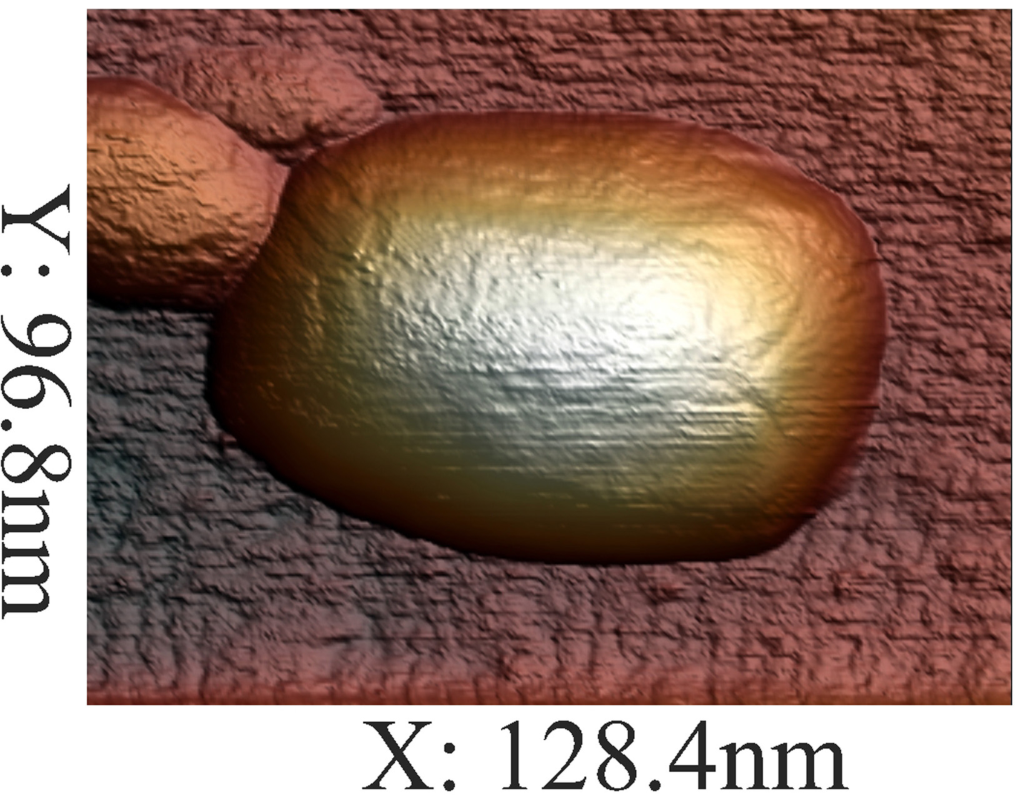}(c)
\includegraphics[scale=1]{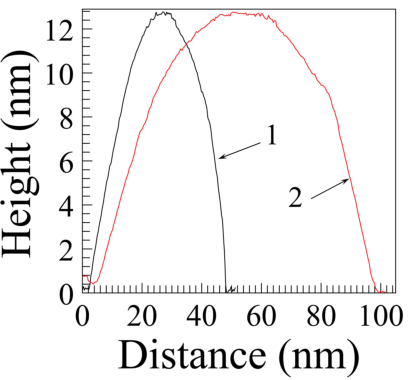}(d)
\end{minipage}\hspace{1pc}%
\begin{minipage}[b]{11.9pc}
\caption{\label{fig:5min_clasters} 
STM images of Ge clusters forming on the Ge/Si(001) wetting layer ($h_{\rm Ge} = 7\,${\AA}) after Ge deposition at room temperature followed by annealing at 600\,{\textcelsius} for 5 min: (a) a derivative image demonstrating numerous clusters of different sizes distributed over the surface; (b,\,c) typical  oval clusters; (d) profiles of the oval cluster shown in panel (c) across (1) and along (2) the cluster (height from the wetting layer vs distance along the wetting layer).
 }
\end{minipage}
\end{figure}  

\subsection{\label{sec:Ge_deposition}Ge film deposited at room temperature}

Films of Ge deposited at room temperature on the buffer layers of Si are composed of densely packed grains (figure \ref{fig:poly-Ge}).
The sizes of the grains vary from less than nanometer to a few nanometers. 
The films are similar in their structure to the films which are formed at room temperature directly on the clean (001) surface of a silicon substrate \cite{Anneal@600C}. The film is seen to be partially ordered; it is composed by relatively long nearly parallel chains consisting of tens grains.
Terraces and steps of the underlaying Si buffer (both ${\rm S_a}$ and ${\rm S_b}$) are clearly seen in the STM images presented in figure \ref{fig:poly-Ge}: the granular film strictly follows the relief of the underlying surface and being smooth enough does not conceal the details of this minor roughness. This observation allows us to conclude that the obtained granular film does not reconstruct the Si buffer at least on the level of the structure of its steps.

\begin{figure}[t]
\includegraphics[width=12pc]{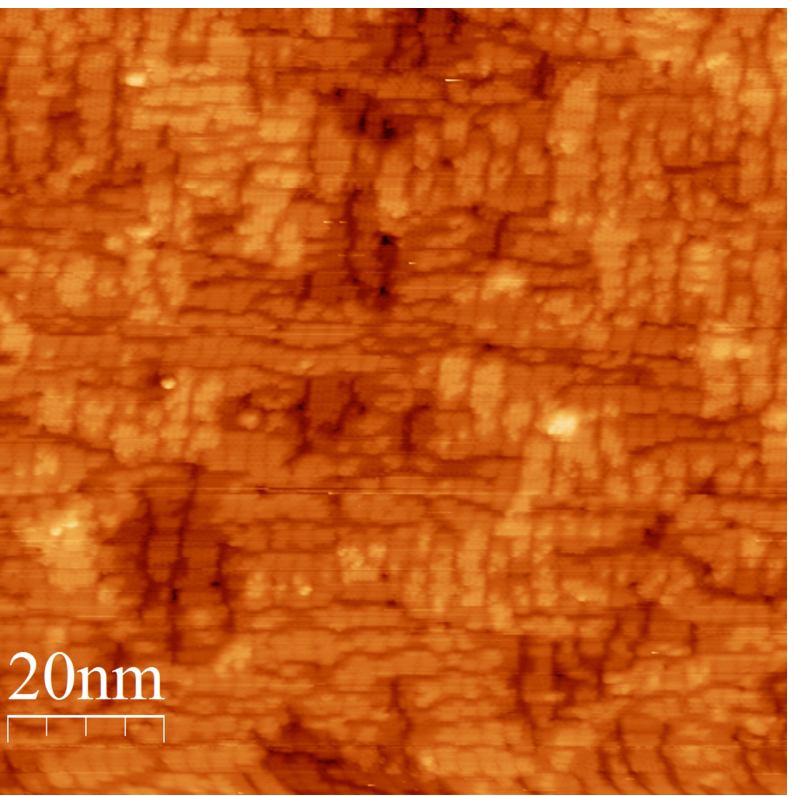}(a)
\includegraphics[width=12pc]{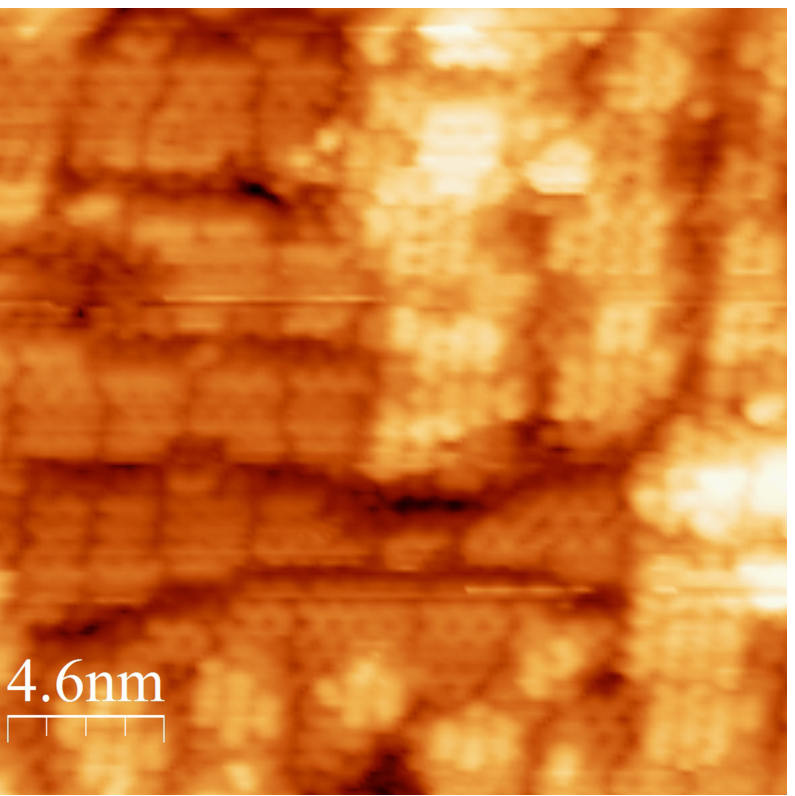}(b)\hspace{1pc}%
\begin{minipage}[b]{10.5pc}
\caption{\label{fig:5min_WL}
STM images of the Ge/Si(001) wetting layer  after Ge deposition at room temperature ($h_{\rm Ge}$ = 7\,{\AA})
followed by annealing at 600\,{\textcelsius} for 5 min: (a) a general view and (b) a mixed reconstruction of patches.
}
\end{minipage}
\end{figure}

\begin{figure}[t]
\includegraphics[width=12pc]{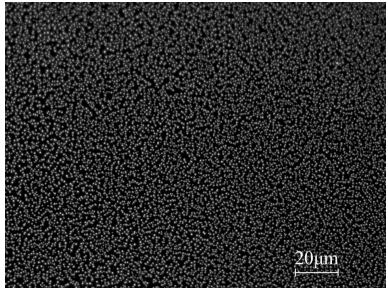}\hspace{2pc}
\begin{minipage}[b]{23.6pc}
\caption{\label{fig:5min_photo}
A microphotograph of the sample surface after Ge deposition at room temperature ($h_{\rm Ge}$ = 7\,{\AA})
followed by annealing at 600\,{\textcelsius} for 5 min; the scale mark is 20\,{\textmu}m.
}
\end{minipage}
\end{figure}

\begin{figure}[t]
\includegraphics[width=12pc]{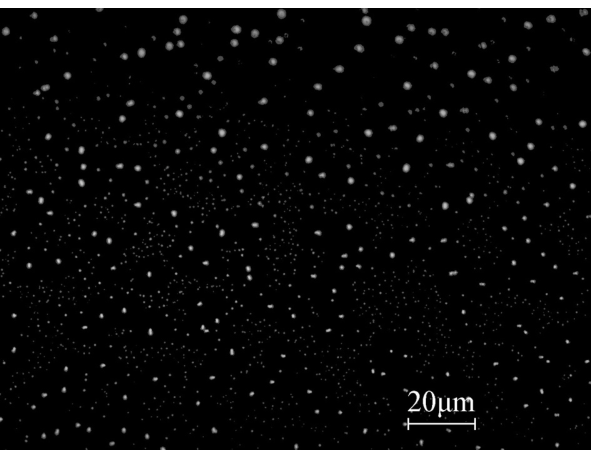}(a)%
\includegraphics[width=12pc]{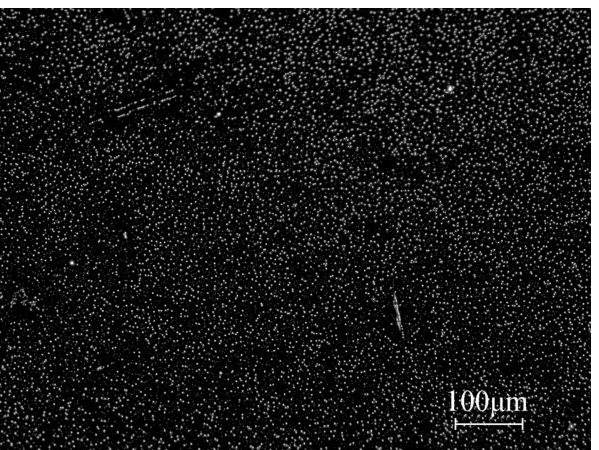}(b)\hspace{1pc}%
\begin{minipage}[b]{10.5pc}
\caption{\label{fig:125min_photo}
Microphotographs of the sample surface after Ge deposition at room temperature ($h_{\rm Ge}$ = 7\,{\AA})
followed by annealing at 600\,{\textcelsius} for 125 min; the scale marks are (a) 20 and (b) 100\,{\textmu}m.
}
\end{minipage}
\end{figure}

\subsection{\label{sec:5min_annealing}Ge layer annealed for 5 minutes at 600\,{\textcelsius}}

After rapid heating and isothermal annealing of the samples at 600\,{\textcelsius} for 5 minutes, i.e. as a result of thermal treatment in the conditions of a closed system and in the conditions of an isolated system for the last 5 minutes,  the deposited at room temperature Ge granular film reconstructs into the usual wetting layer with oval (partially faceted) Ge mounds (figures~\ref{fig:5min_clasters} and~\ref{fig:5min_WL}). The final structure coincides with that observed in the reference \cite{Anneal@600C} after the same thermal processing of similar films deposited directly on the clean (001) surfaces of silicon wafers. The number density of the Ge droplets estimated from the STM images makes (1.5~to~$2)\times 10^9\,$cm$^{-2}$ that slightly exceeds the estimates obtained in the reference \cite{Anneal@600C} (figure~\ref{fig:5min_clasters}). The drops sizes usually exceed 100\,nm in lateral dimensions and reach 15\,nm in height. Note that there are many coalescent clusters observed on the surface; in addition, small and relatively large drops are observed simultaneously.

The wetting layer structure resembles one typical for the wetting layers formed at low temperatures (around 350\textcelsius) during MBE \cite{initial_phase,Yur_JNO,SPIE_QD-chains,Growing_Ge_hut-structure}. Tops  of the patches are composed by $c(4\times 2)$ and $p(2\times 2)$ reconstructed layers in close proportions (figure \ref{fig:5min_WL}) \cite{initial_phase,VCIAN2011}. This mixed type of the wetting layer reconstruction is likely a result of a kinetically controlled transformation.

In addition to the oval drops large clusters of Ge were also detected on the surface using a light microscope (figure~\ref{fig:5min_photo}). Since their number density estimated from the microphotographs was about $5\times 10^7\,$cm$^{-2}$ a probability of their detection with STM was very low. Unfortunately, now we cannot judge  about their shapes; their sizes can be estimated as at least several hundreds of nanometers. Their size distribution is at least bimodal; they often form chains.

\subsection{\label{sec:125min_annealing}Ge layer annealed for 125 minutes at 600\,{\textcelsius}} 

After rapid heating and isothermal annealing of the samples at 600\,{\textcelsius} for 125 minutes, the large clusters of Ge increase their sizes but their distribution remains multimodal (figure \ref{fig:125min_photo}); the number density of these clusters reduces by two orders of magnitude to $\sim 5\times 10^5\,$cm$^{-2}$ (figure~\ref{fig:125min_photo}a) however they retain some tendency to form chains (figure \ref{fig:125min_photo}b). The largest clusters are seen to consume the substance from the surrounding smaller ones and form an empty space around themselves (figure~\ref{fig:125min_photo}a).

\begin{figure}[t]
\centering
\includegraphics[scale=.55]{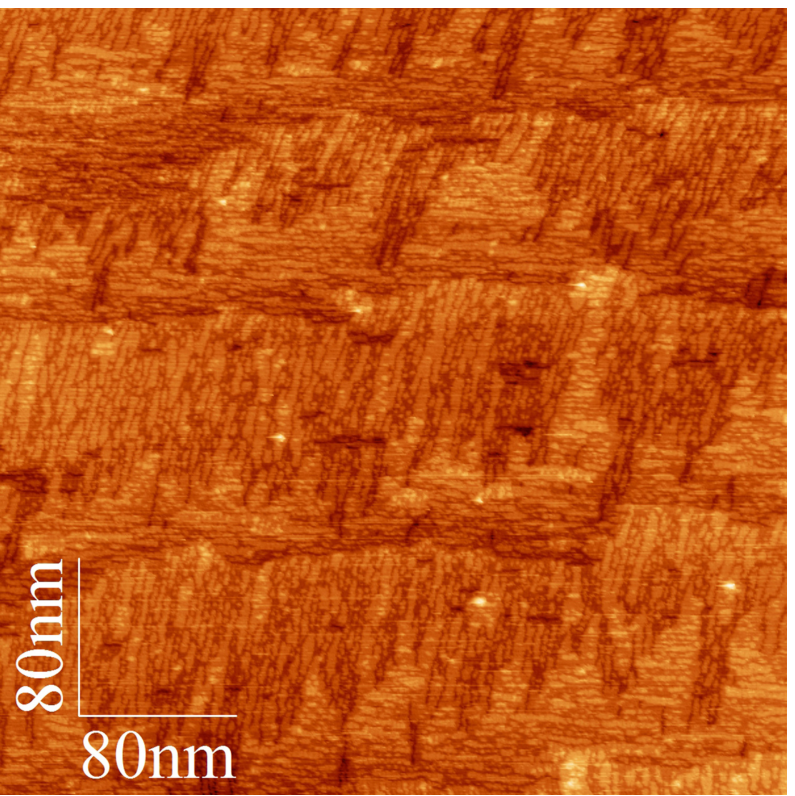}(a)
\includegraphics[scale=.55]{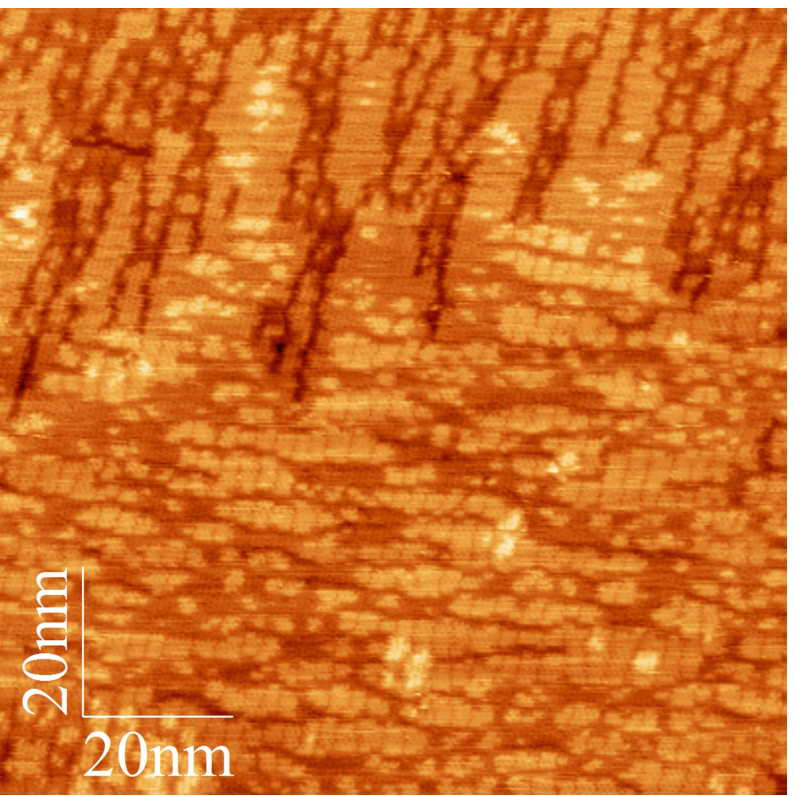}(b)
\includegraphics[scale=.55]{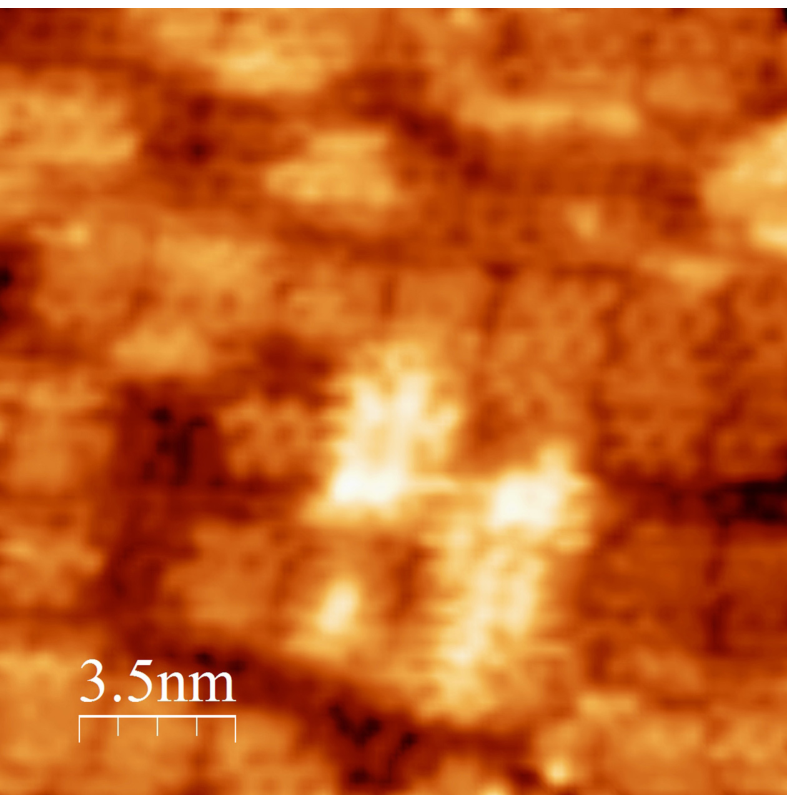}(c)
\caption{\label{fig:125min}
STM images of the Ge/Si(001) layer ($h_{\rm Ge}$ = 7\,{\AA}) after Ge deposition at room temperature followed by annealing at 600\,{\textcelsius} for 125 min.
}
\end{figure}

STM images do not show any presence of the oval drops on the wetting layer (figure \ref{fig:125min}). At the same time, no pyramids or domes are also observed on the surface. And evidently, wedges are not observed, either \cite{Nucleation_high-temperatures}. 
Patches of the Ge wetting layer are mainly $c(4\times 2)$-reconstructed. This type of the wetting layer reconstruction is characteristic to the high-temperature mode   of the MBE growth of Ge on Si(001) ($T_{\rm gr}>600\,$\textcelsius) \cite{Nucleation_high-temperatures}. We conclude that the $c(4\times 2)$ reconstruction forms under the thermodynamic control and is thermodynamically favoured.

We believe that the observed disappearance of the Ge drops from the wetting layer after sample annealing  at 600\,{\textcelsius} for 125 minutes is explained by Ge redistribution in favour of the growing large clusters.
Ge and Si intermixing may also affect the drop disappearance.

\section{\label{sec:conclusion}Conclusion}
In summary, as a result of the short-term annealing at 600\,{\textcelsius} at the described above conditions of an isolated system the Ge/Si(001) granular film deposited at room temperature transforms to the usual wetting layer and Ge clusters with multimodal size distribution with the Ge oval drops having the highest number density. After the long-term thermal treatment of the described initial samples at the same temperature, Ge drops disappear; the large clusters increase their sizes at the expense of the smaller ones. The total density of Ge clusters on the surface decreases by about four orders of magnitude. The wetting layer mixed $c(4\times 2)+p(2\times 2)$ reconstruction forming after the sort-term annealing transforms to the single $c(4\times 2)$ one, which is likely thermodynamically favoured, as a result of the long-term annealing. Pyramids or domes are not observed on the surface after any annealing.



\ack
We cordially thank Ms N~V~Kiryanova for her invaluable contribution to arrangement of this research. We express our appreciation to Mr O~V~Uvarov for obtaining microphotographs. 
We also thank Ms L~M~Krylova for chemical treatments of the samples.
The Center of Collective Use of Scientific Equipment of GPI RAS supported this research by presenting admittance to its instrumentation. We acknowledge the support of this research.

\section*{References}



\providecommand{\newblock}{}

\end{document}